\documentclass{jps-cp}
\usepackage{txfonts} %Please comment out this line unless the txfonts package is availabe in your LaTeX system.
\usepackage{multirow}
\usepackage{bm, amsmath, amsfonts, amssymb, braket}
\usepackage{times}
\usepackage{multirow}
\usepackage[dvipdfmx]{graphicx}
\usepackage{float}
\usepackage{color}
\usepackage{comment}
\usepackage{cite}
\usepackage{nccmath} % &=& space

\newcommand{\ii}{\text{i}}
\newcommand{\Z}{$\mathbb{Z}$}
\newcommand{\Zt}{$\mathbb{Z}_{2}$}
\newcommand{\sgn}{$\text{sgn}$}

\newcommand{\Lr}{~$\text{L}_{\rm r}$}
\newcommand{\Li}{~$\text{L}_{\rm i}$}

\newcommand{\Ca}{${\cal C}_{p}$ & \Z & $0$ & \Z & $0$}
\newcommand{\Caa}{${\cal C}_{p} \times {\cal C}_{p}$ & $\mathbb{Z} \oplus \mathbb{Z}$ & $0$ & $\mathbb{Z} \oplus \mathbb{Z}$ & $0$}

\newcommand{\Cb}{${\cal C}_{p+1}$ & $0$ & \Z & $0$ & \Z }
\newcommand{\Cbb}{${\cal C}_{p+1} \times {\cal C}_{p+1}$ & $0$ & $\mathbb{Z} \oplus \mathbb{Z}$ & $0$ & $\mathbb{Z} \oplus \mathbb{Z}$ }

\newcommand{\Ra}{${\cal R}_{p}$ & \Z & \Zt & \Zt & $0$}

\newcommand{\Rb}{${\cal R}_{p+1}$ & \Zt & \Zt & $0$ & 2\Z }
\newcommand{\Rbb}{${\cal R}_{p+1} \times {\cal R}_{p+1}$ & $\mathbb{Z}_{2} \oplus \mathbb{Z}_{2}$ & $\mathbb{Z}_{2} \oplus \mathbb{Z}_{2}$ & $0$ & $2\mathbb{Z} \oplus 2\mathbb{Z}$ }

\newcommand{\Rc}{${\cal R}_{p+2}$ & \Zt & $0$ & 2\Z & $0$}

\newcommand{\Rd}{${\cal R}_{p+3}$ & $0$ & 2\Z & $0$ & $0$}
\newcommand{\Rdd}{${\cal R}_{p+3} \times {\cal R}_{p+3}$ & $0$ & $2\mathbb{Z} \oplus 2\mathbb{Z}$ & $0$ & $0$}

\newcommand{\Ree}{${\cal R}_{p+4}$ & 2\Z & $0$ & $0$ & $0$ }

\newcommand{\Rf}{${\cal R}_{p+5}$ & $0$ & $0$ & $0$ & \Z }
\newcommand{\Rff}{${\cal R}_{p+5} \times {\cal R}_{p+5}$ & $0$ & $0$ & $0$ & $\mathbb{Z} \oplus \mathbb{Z}$ }

\newcommand{\Rg}{${\cal R}_{p+6}$ & $0$ & $0$ & \Z & \Zt }

\newcommand{\Rh}{${\cal R}_{p+7}$ & $0$ & \Z & \Zt & \Zt & }
\newcommand{\Rhh}{${\cal R}_{p+7} \times {\cal R}_{p+7}$ & $0$ & $\mathbb{Z} \oplus \mathbb{Z}$ & $\mathbb{Z}_{2} \oplus \mathbb{Z}_{2}$ & $\mathbb{Z}_{2} \oplus \mathbb{Z}_{2}$}

\newcommand{\pmat}[1]{\begin{pmatrix} #1 \end{pmatrix}}%matrix

\title{Topological Classificaton of Non-Hermitian Gapless Phases: Exceptional Points and Bulk Fermi Arcs}

\author{Takumi \textsc{Bessho}$^{1}$, Kohei \textsc{Kawabata}$^{2}$, and Masatoshi \textsc{Sato}$^{1}$}

\inst{$^{1}$Yukawa Institute for Theoretical Physics, Kyoto University, Kyoto 606-8502, Japan \\
$^{2}$ Department of Physics, University of Tokyo, 7-3-1 Hongo, Bunkyo-ku, Tokyo 113-0033, Japan}

\email{takumi.bessho@yukawa.kyoto-u.ac.jp}

\recdate{September 17, 2019}

\abst{
We provide classification of gapless phases in non-Hermitian systems according to two types of complex-energy gaps: point gap and line gap. 
We show that exceptional points, at which not only eigenenergies but also eigenstates coalesce, are characterized by gap closing of point gaps with nontrivial topological charges.
Moreover, we find that bulk Fermi arcs accompanying exceptional points are robust because of topological charges for real line gaps.
On the basis of the classification, some examples are also discussed.
}

\kword{non-Hermitian, topological, exceptional point, symmetry}

\begin{document}
\maketitle

\section{Introduction}

Recently, there has been much interest in non-Hermitian physics \cite{Bender-98, Hatano}.
PT-symmetry breaking, which is spontaneous symmetry breaking that accompanies an exceptional point, was experimentally observed in photonic systems with gain and/or loss \cite{Ruter-10}. It was revealed that exceptional points lead to phenomena unique to non-Hermitian systems, such as unidirectional invisibility and enhanced sensitivity \cite{Ganainy-18}.
It was also pointed out that non-Hermitian Hamiltonians are relevant to many-body systems with finite-lifetime quasiparticles\cite{Kozii-18}. There, a bulk Fermi arc appears between a pair of exceptional points, which is expected to increase the density of states and change many physical quantities of solids. 
Moreover, exceptional points, lines, and surfaces protected by PT \cite{Okugawa-19}, CP, and chiral symmetries \cite{Budich-19, Yoshida-19, Kimura-19} were studied.
However, classification theory that provides a unified understanding about such a variety of non-Hermitian gapless structures has yet to be established.

In this paper, we outline a unified theory that classifies exceptional points in a general manner and predicts unknown symmetry-protected exceptional points \cite{KBS-19}.
We topologically classify non-Hermitian gapless phases according to two types of complex-energy gaps, point gap and line gap, which are defined in Sec. 2.
Exceptional points ($E=0$) are characterized as point-gapless points with nontrivial topological charges (Tables I-III, P gap).
On the other hand, bulk Fermi arcs ($\text{Re}E = 0$) accompanying exceptional points are characterized as real-line-gapless points (Tables I-III, L gap). Multiple topological structures of non-Hermitian semimetals are understood by our classification, as illustrated with examples in Sec. 6

\section{Gap structure of exceptional points}
In non-Hermitian systems, energy spectra become complex in general. Hence there are two natural definitions of complex-energy gaps: point gap and line gap (Fig.~1)\cite{KSUS-19}. 
In the presence of a point (line) gap, complex-energy bands do not cross a reference point (line) in the complex energy plane.
This is a generalization of the energy gap in Hermitian systems where the energy spectra are real. In the Hermitian case, we usually take an energy gap at $E=0$ because of symmetries.
For example, particle-hole symmetry leads to $(E, -E)$ pairs and hence the energy gap should be taken as $E=0$ to respect the symmetry.
Similarly, in the non-Hermitian case, we take a point gap at $E=0$. 
We also take line gaps at Re$E=0$ and Im$E=0$. We note that point and line gaps can be taken arbitrarily in the absence of symmetries. Remarkably, exceptional points are described by point-gapless points ($E=0$), while bulk Fermi arcs are described by real-line-gapless points ($\text{Re} E = 0$).

\begin{figure}[tbh]
 \begin{center}
\includegraphics[width=100mm]{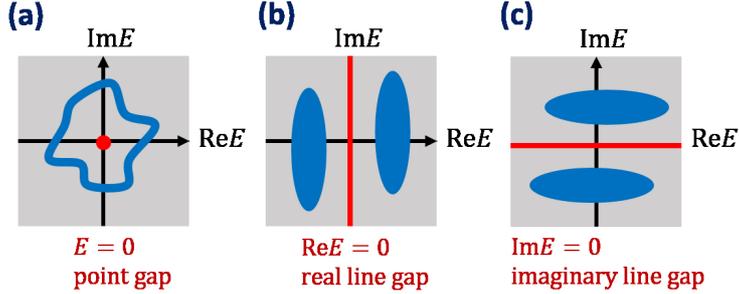}
\caption{(a) Point gap, (b) real line gap, and (c) imaginary line gap in the complex energy plane. A blue curve and regions are energy band spectra. The red point and lines are a reference point of a point gap and reference lines of line gaps, respectively.
}
 \end{center}
\label{fig:gap}
\end{figure}

We define an exceptional point as a defective point in this work. Let us consider the following Hamiltonian as an example:
\begin{eqnarray}
H=\pmat{0 & \epsilon +k_x -ik_y \\
k_x+ik_y & 0},
\end{eqnarray}
where $\epsilon >0 $ represents a degree of non-Hermiticity.
This Hamiltonian is defective at $\bm{k}_{\text{EP}}=(0,0)$, and thus has only one eigenstate $\ket{\psi (0,0)}=(1,0)$ and eigenenergy $E(0,0)=0$. In general, if an $n\times n$ matrix has fewer than $n$ independent eigenvectors, we call the matrix non-diagonalizable or defective.

The energy spectrum near the exceptional point $\bm{k}_{\rm EP}$ becomes $E(\bm{k})\approx \pm \sqrt{\epsilon (k_x+\ii k_y)}$.
The energy spectrum exhibits a square-root singularity for ${\bm k}$ around $\bm{k}_{\rm EP}$, which is impossible in Hermitian systems.
Non-analyticity indicates that the conventional perturbation theory does not hold at exceptional points.

From the perspective of gap structures, we find that exceptional points are nontrivial point-gapless points.
For Eq.~(1), if we take a circle around the exceptional point $S^1:  k_x^2+k_y^2=r^2$ with a radius $0<r<\epsilon$, a point gap is open on this circle and hence a topological invariant (winding number for complex eigenenergies \cite{Leykam-17, Shen-18}) is obtained as:
\begin{eqnarray}\label{eq:winding}
W :=\oint_{S^{1}} \frac{d \bm{k}}{2 \pi \mathrm{i}} \cdot \nabla_{\bm{k}} \log \operatorname{det} H(\bm{k})
=\int_0^{2\pi} \frac{d \theta}{2 \pi \mathrm{i}} \partial_\theta \log [-(\epsilon+r e^{-\ii \theta})re^{\ii \theta}]=1.
\end{eqnarray}
Thus the exceptional point is a nontrivial point-gapless point with the topological charge $W=1$, which means the topological stability of the exceptional point.
This result indicates that exceptional points are topologically classified as point-gapless points.

\section{Symmetries}
In this work, we consider symmetries that keep each momentum invariant.
Such symmetries protect exceptional points at generic momentum points and are included in $\mathcal{P}\text{AZ}$ symmetry, $\mathcal{P}\text{AZ}^{\dag}$ symmetry, and additional symmetries such as pseudo-Hermiticity and sublattice symmetry \cite{KSUS-19, KBS-19}.

${\cal P}$AZ symmetry consists of $\mathcal{P}\mathcal{T}_{+}$ symmetry, $\mathcal{C}_{-}\mathcal{P}$ symmetry, and chiral symmetry: 
\begin{eqnarray}
({\cal P}{\cal T}_{+})\,H^{*} \left( {\bm k} \right) ({\cal P}{\cal T}_{+})^{-1} &=& H \left( {\bm k} \right), \quad
 ({\cal P}{\cal T}_{+}) ({\cal P}{\cal T}_{+})^{*} = \pm 1.
 \\
 ({\cal C}_{-}{\cal P})\,H^{T} \left( {\bm k} \right) ({\cal C}_{-}{\cal P})^{-1} &=& - H \left( {\bm k} \right), \quad
 ({\cal C}_{-}{\cal P}) ({\cal C}_{-}{\cal P})^{*} = \pm 1.
 \\
 \Gamma\,H^{\dag} \left( {\bm k} \right) \Gamma^{-1} &=& - H \left( {\bm k} \right), \quad
 \Gamma^{2} = 1.
\end{eqnarray}
Here, ${\cal P}{\cal T}_{+}$, ${\cal C}_{-}{\cal P}$, and $\Gamma$ are unitary operators.
 $\mathcal{P}\mathcal{T}_{+}$ symmetry makes the energy spectrum real even in non-Hermitian systems if the eigenstates also respect ${\cal P}{\cal T}_{+}$ symmetry \cite{Bender-98}. 
 $\mathcal{C}_{-}\mathcal{P}$ symmetry is relevant to non-Hermitian Bogoliubov-de Gennes Hamiltonians with inversion symmetry.

${\cal P} \text{AZ}^{\dag}$ symmetry consists of $\mathcal{C}_{+}\mathcal{P}$ symmetry, $\mathcal{P}\mathcal{T}_{-}$ symmetry, and chiral symmetry: 
\begin{eqnarray}
({\cal C}_{+}{\cal P})\,H^{T} \left( {\bm k} \right) ({\cal C}_{+}{\cal P})^{-1} &=& H \left( {\bm k} \right), \quad
({\cal C}_{+}{\cal P}) ({\cal C}_{+}{\cal P})^{*} = \pm 1.
\\
({\cal P}{\cal T}_{-})\,H^{*} \left( {\bm k} \right) ({\cal P}{\cal T}_{-})^{-1} &=& - H \left( {\bm k} \right), \quad
({\cal P}{\cal T}_{-}) ({\cal P}{\cal T}_{-})^{*} = \pm 1.
\end{eqnarray}
Here, ${\cal C}_{+}{\cal P}$ and ${\cal P}{\cal T}_{-}$ are unitary operators.
Pseudo-Hermiticity and sublattice symmetry are
\begin{eqnarray}
\eta H^{\dag} \left( \bm k \right) \eta^{-1} &=& H \left( \bm k \right),\quad \eta^2=1.\\
{\cal S} H \left( {\bm k} \right) {\cal S}^{-1} &=& - H \left( {\bm k} \right),\quad
{\cal S}^{2} = 1.
\end{eqnarray}
Here, $\eta$ and ${\cal S}$ are unitary operators.
$\mathcal{C}_{+}\mathcal{P}$ symmetry and $\mathcal{P}\mathcal{T}_{-}$ symmetry are equivalent to $\mathcal{P}\mathcal{T}_{+}$ symmetry and $\mathcal{C}_{-}\mathcal{P}$ symmetry for Hermitian Hamiltonians, respectively. However, this is not the case for non-Hermitian Hamiltonians because of $H^{*} \neq H^{T}$.
Combining all the above symmetries, we obtain 38 symmetry classes in total as a non-Hermitian generalization of the 10-fold Altland-Zirnbauer (AZ) symmetry class~\cite{KSUS-19, KBS-19}. Below, we give topological classification tables for 17 independent symmetry classes of them.

\section{Topological classification of non-Hermitian gapless phases}
We topologically classify the gapless points, lines, and surfaces.
In Hermitian systems, a Weyl point in three dimensions is characterized by the Chern number defined on a sphere $S^2$ surrounding the Weyl point.
Similarly, topology of non-Hermitian gapless systems reduces to the gapped counterpart.
In fact, a gapless point in $p$ dimension is characterized by a topological invariant defined on a $(p-1)$-dimensional sphere $S^{p-1}$ that encloses that gapless point even in non-Hermitian systems.

Furthermore, non-Hermitian point/line-gapped phases reduce to Hermitian gapped phases.
A non-Hermitian point-gapped Hamiltonian $H \left( {\bm k} \right)$ has the following correspondence with a Hermitian gapped Hamiltonian $\bar{H} \left( {\bm k} \right)$:
\begin{eqnarray}
\bar{H}(\bm{k}) :=\pmat{ & H(\bm{k}) \\ H^\dagger (\bm{k}) & }, \quad
 \Sigma_{z} \bar{H}(\bm{k}) \Sigma_{z}=-\bar{H}(\bm{k}), \quad
 \Sigma_{z} :=\left(\begin{array}{cc}{I} & {} \\ {} & {-I}\end{array}\right).
\end{eqnarray}
Notably, additional chiral (sublattice) symmetry described by $\Sigma_z$ is respected for $\bar{H} \left( {\bm k} \right)$ by construction, which leads to unique non-Hermitian topology for point gaps.
On the other hand, a non-Hermitian real-line-gapped Hamiltonian is continuously deformed into a Hermitian gapped Hamiltonian without closing the real line gap. 
A non-Hermitian imaginary-line-gapped Hamiltonian is mapped into a non-Hermitian real-line-gapped Hamiltonian, multiplying it by the imaginary unit: $H \to \ii H$. Then, we can reduce it into a Hermitian Hamiltonian similarly.
Detailed proofs for the above complex-spectral-flattening procedures are described in Ref.~\cite{KSUS-19}.

\section{Classification tables}

Tables I-III are topological classification tables for the  17 independent symmetry classes and the two types of energy gaps. Note that class AIII, ${\cal P}$AI, and ${\cal P}$AII are related to pH, ${\cal P}$D$^{\dagger}$, and ${\cal P}$C$^{\dagger}$ by the transformation $H \to \text{i}H$.
All the classification tables of 38 symmetry classes are found in Ref.~\cite{KBS-19}. 

%%%%% AZ %%%%%
\begin{table}[htbp]
	\centering
	\caption{Topological classification table of non-Hermitian gapless phases in ${\cal P}$AZ symmetry class. Non-Hermitian gapless phases are classified according to codimension $p$, ${\cal P}$AZ symmetry class, and two types of complex-energy gaps [point (P) and line (L) gaps].
Codimension $p$ is defined as $p:=d-d_G$ with spatial dimension $d$ and the dimension $d_G$ of the gapless points ($d_G=0$), lines ($d_G=1$), surfaces ($d_G=2$), and volumes ($d_G=3$). \Lr~and \Li~indicate a real line gap and an imaginary line gap, respectively. \\}
	\label{tab: complex + real AZ}
     \begin{tabular}{ccccccccccc} \hline \hline
    ~${\cal P}\text{AZ}$ class~ & $\mathcal{P}\mathcal{T}_+$ & $\mathcal{C}_-\mathcal{P}$ & $\Gamma$ & ~Gap~ & Classifying space & ~~$p=0$~~ & ~$p=1$~ & ~$p=2$~ & ~$p=3$~ \\ \hline
    \multirow{2}{*}{A}
    & \multirow{2}{*}{0} & \multirow{2}{*}{0} & \multirow{2}{*}{0}
    & P & \Ca \\ 
    &&&& L & \Cb \\ \hline
    \multirow{3}{*}{AIII} 
    & \multirow{3}{*}{0} & \multirow{3}{*}{0} & \multirow{3}{*}{1}
    & P & \Cb \\ 
    &&&& \Lr & \Ca \\
    &&&& \Li & \Cbb \\ \hline \hline
    \multirow{3}{*}{${\cal P}$AI} 
    & \multirow{3}{*}{+1} & \multirow{3}{*}{0} & \multirow{3}{*}{0}
    & P & \Ra \\ 
    &&&& \Lr & \Rh\\
    &&&& \Li & \Rb \\ \hline
    \multirow{3}{*}{${\cal P}$BDI} 
    & \multirow{3}{*}{+1} & \multirow{3}{*}{+1} & \multirow{3}{*}{1}
    & P & \Rb \\ 
    &&&& \Lr & \Ra \\
    &&&& \Li & \Rbb \\ \hline
    \multirow{2}{*}{${\cal P}$D} 
    & \multirow{2}{*}{0} & \multirow{2}{*}{+1} & \multirow{2}{*}{0}
    & P & \Rc \\ 
    &&&& L & \Rb \\ \hline
    \multirow{3}{*}{${\cal P}$DIII} 
    & \multirow{3}{*}{-1} & \multirow{3}{*}{+1} & \multirow{3}{*}{1}
    & P & \Rd \\ 
    &&&& \Lr & \Rc \\
    &&&& \Li & \Cb \\ \hline
    \multirow{3}{*}{${\cal P}$AII} 
    & \multirow{3}{*}{-1} & \multirow{3}{*}{0} & \multirow{3}{*}{0}
    & P & \Ree \\ 
    &&&& \Lr & \Rd \\
    &&&& \Li & \Rf \\ \hline
    \multirow{3}{*}{${\cal P}$CII} 
    & \multirow{3}{*}{-1} & \multirow{3}{*}{-1} & \multirow{3}{*}{1}
    & P & \Rf \\ 
    &&&& \Lr & \Ree \\
    &&&& \Li & \Rff \\ \hline
    \multirow{2}{*}{${\cal P}$C} 
    & \multirow{2}{*}{0} & \multirow{2}{*}{-1} & \multirow{2}{*}{0}
    & P & \Rg \\ 
    &&&& L & \Rf \\ \hline
    \multirow{3}{*}{${\cal P}$CI} 
    & \multirow{3}{*}{+1} & \multirow{3}{*}{-1} & \multirow{3}{*}{1}
    & P & \Rh \\ 
    &&&& \Lr & \Rg \\
    &&&& \Li & \Cb \\ \hline \hline
  \end{tabular}
\end{table}
%%%%% classification table %%%%%
\begin{table*}[htbp]
	\centering
	\caption{Topological classification table of non-Hermitian gapless phases in the presence of pseudo-Hermiticity or sublattice symmetry. \\}
	\label{tab: classification}
     \begin{tabular}{ccccccccccc} \hline \hline
    ~~Symmetry~~ & ~Gap~ & ~Classifying space~ & ~~$p=0$~~ & ~~$p=1$~~ & ~~$p=2$~~ & ~~$p=3$~~ \\ \hline
    \multirow{3}{*}{pH} 
    & P & \Cb \\ 
    & \Lr & \Cbb \\
    & \Li & \Ca \\ \hline
    \multirow{2}{*}{Sublattice} 
    & P & \Caa \\ 
    & L & \Ca \\ \hline \hline
  \end{tabular}
\end{table*}
%%%%%%%%%%

\clearpage

%%%%% real AZ^{\dag} %%%%%
\begin{table}[t]
	\centering
	\caption{Topological classification table of non-Hermitian gapless phases in ${\cal P}$AZ$^\dagger$ symmetry class. \\}
     \begin{tabular}{ccccccccccc} \hline \hline
    ~${\cal P}\text{AZ}^{\dag}$ class~ & $\mathcal{C}_+\mathcal{P}$ & $\mathcal{P}\mathcal{T}_-$ & $\Gamma$ & ~Gap~ & Classifying space  & ~~$p=0$~~ & ~$p=1$~ & ~$p=2$~ & ~$p=3$~ \\ \hline
   \multirow{2}{*}{A}
    & \multirow{2}{*}{0} & \multirow{2}{*}{0} & \multirow{2}{*}{0}
    & P & \Ca \\ 
    &&&& L & \Cb \\ \hline
    \multirow{3}{*}{AIII} 
    & \multirow{3}{*}{0} & \multirow{3}{*}{0} & \multirow{3}{*}{1}
    & P & \Cb \\ 
    &&&& \Lr & \Ca \\
    &&&& \Li & \Cbb \\ \hline \hline
    \multirow{2}{*}{${\cal P}\text{AI}^{\dag}$} 
    & \multirow{2}{*}{+1} & \multirow{2}{*}{0} & \multirow{2}{*}{0}
    & P & \Rg \\ 
    &&&& L & \Rh \\ \hline
    \multirow{3}{*}{${\cal P}\text{BDI}^{\dag}$} 
    & \multirow{3}{*}{+1} & \multirow{3}{*}{+1} & \multirow{3}{*}{1}
    & P & \Rh \\ 
    &&&& \Lr & \Ra \\
    &&&& \Li & \Rhh \\ \hline
    \multirow{3}{*}{${\cal P}\text{D}^{\dag}$} 
    & \multirow{3}{*}{0} & \multirow{3}{*}{+1} & \multirow{3}{*}{0}
    & P & \Ra \\
    &&&& \Lr & \Rb \\ 
    &&&& \Li & \Rh \\ \hline
    \multirow{3}{*}{${\cal P}\text{DIII}^{\dag}$} 
    & \multirow{3}{*}{-1} & \multirow{3}{*}{+1} & \multirow{3}{*}{1}
    & P & \Rb \\ 
    &&&& \Lr & \Rc \\
    &&&& \Li & \Cb \\ \hline
    \multirow{2}{*}{${\cal P}\text{AII}^{\dag}$} 
    & \multirow{2}{*}{-1} & \multirow{2}{*}{0} & \multirow{2}{*}{0}
    & P & \Rc \\ 
    &&&& L & \Rd \\ \hline
    \multirow{3}{*}{${\cal P}\text{CII}^{\dag}$} 
    & \multirow{3}{*}{-1} & \multirow{3}{*}{-1} & \multirow{3}{*}{1}
    & P & \Rd \\ 
    &&&& \Lr & \Ree \\
    &&&& \Li & \Rdd \\ \hline
    \multirow{3}{*}{${\cal P}\text{C}^{\dag}$} 
    & \multirow{3}{*}{0} & \multirow{3}{*}{-1} & \multirow{3}{*}{0}
    & P & \Ree \\ 
    &&&& \Lr & \Rf \\
    &&&& \Li & \Rd \\ \hline
    \multirow{3}{*}{${\cal P}\text{CI}^{\dag}$} 
    & \multirow{3}{*}{+1} & \multirow{3}{*}{-1} & \multirow{3}{*}{1}
    & P & \Rf \\ 
    &&&& \Lr & \Rg \\
    &&&& \Li & \Cb \\ \hline \hline
  \end{tabular}
\end{table}
%%%%%%%%%%

\section{Examples}
We give some examples to illustrate that both exceptional points ($E=0$) and bulk Fermi arcs ($\text{Re}E=0$) are topologically stable according to the classification tables (Tables I-III).
Importantly, non-Hermitian semimetals can possess multiple topological structures due to two types of complex-energy gaps. We first consider a one-dimensional model with exceptional points in class AIII (Table I):
\begin{eqnarray}
H(k)=k \sigma_x +\ii \gamma \sigma_z, \quad 
E(k)=\pm \sqrt{k^2-\gamma^2},
\end{eqnarray}
where $\gamma \in \mathbb{R}$ represents a degree of non-Hermiticity.
This Hamiltonian has chiral symmetry $\Gamma\,H^{\dag} \left( {\bm k} \right) \Gamma^{-1} = - H \left( {\bm k} \right)$ with $\Gamma = \sigma_z$, and has a pair of exceptional points at $k_{\text{EP}}=\pm \gamma$. These exceptional points have nontrivial $\mathbb{Z}$ topology for a point gap (see Table I with class AIII and codimension $p=1$). The corresponding topological invariant is the difference of the occupation numbers (zeroth Chern numbers) of the Hermitian matrix $\ii H(k) \Gamma=k\sigma_y-\gamma$ defined at a pair of momenta across the exceptional point, which is $\pm 1$ for $k_{\text{EP}}=\pm \gamma$, respectively.
The exceptional points are also accompanied by a Fermi arc with $\mathrm{Re} E = 0$ at $-\gamma < k < \gamma$.
This bulk Fermi arc is robust because of the nontrivial $\mathbb{Z}$ topology for a real line gap (see Table I with class AIII and codimension $p=0$).
The corresponding topological invariant is $\sgn [\braket{\psi |\Gamma| \psi}]$ with a right eigenstate $\ket{\psi}$, and the two eigenstates of Eq.~(11) at $-\gamma < k < \gamma$ have $\sgn [\braket{\psi |\Gamma| \psi}]=\pm 1$.
Similarly to our one-dimensional model, exceptional rings and accompanying Fermi surfaces appear in two dimensions \cite{Yoshida-19}. In three dimensions, exceptional surfaces and accompanying Fermi volumes appear \cite{Kimura-19}. Specific models of many-body systems that support these non-Hermitian gapless phases are also proposed in Refs.~\cite{Yoshida-19,Kimura-19}, which show the enhancement of the specific heat and the magnetic susceptibility due to exceptional points and accompanying Fermi arcs.

We next consider a three-dimensional model with exceptional points in class AIII:
\begin{eqnarray}
H(\boldsymbol{k})=k_{x} \sigma_{x} \tau_{x}+k_{y} \sigma_{x} \tau_{y}+k_{z} \sigma_{x} \tau_{z}+\mathrm{i} \gamma \sigma_{z} \tau_{x}, \quad
E(\bm{k})=\pm \sqrt{k_{x}^{2}+(\sqrt{k_{y}^{2}+k_{z}^{2}} \pm \mathrm{i} \gamma)^{2}}.
\end{eqnarray}
This Hamiltonian has chiral symmetry  $\Gamma\,H^{\dag} \left( {\bm k} \right) \Gamma^{-1} = - H \left( {\bm k} \right)$ with $ \Gamma = \sigma_z$, and has a pair of exceptional points at $\bm{k}_{\text{EP}}=(\pm \gamma, 0,0)$.
These exceptional points have nontrivial $\mathbb{Z}$ topology for a point gap (see Table I with class AIII and codimensions $p=3$). The corresponding topological invariant is the Chern number of the Hermitian matrix $\ii H(\bm{k}) \Gamma$, which is $\pm 1$ for $\bm{k}_{\text{EP}}=(\pm \gamma, 0,0)$, respectively.
The exceptional points are also accompanied by a Fermi arc with $\mathrm{Re} E = 0$ at $-\gamma < k_x < \gamma, \ k_y=k_z=0$.
This bulk Fermi arc is robust because of the nontrivial $\mathbb{Z}$ topology for a real line gap (see Table I with class AIII and codimensions $p=2$).

\section{Summary} 
We provide classification of non-Hermitian gapless phases according to two types of complex-energy gaps: point gap and line gap.
We show that exceptional points are characterized as nontrivial point-gapless phases while bulk Fermi arcs accompanying exceptional points are characterized by real-line-gapless phases.
Such multiple topological structures of non-Hermitian semimetals are understood by the classification tables.

\section{Acknowledgements}
We thank K. Shiozaki for discussions and comments on this work.
This work was supported by a Grant-in-Aid for Scientific Research
on Innovative Areas “\!Topological Materials Science\!” (KAKENHI Grant No. JP15H05855) from the Japan Society for the Promotion of Science (JSPS), and JST CREST (No. JPMJCR19T2), Japan.
K.K. was supported by KAKENHI Grant No.~JP19J21927 from the JSPS.
M.S. was supported by KAKENHI Grant No. JP17H02922 from the JSPS.

\end{document}